\documentclass[10pt]{article}

\usepackage{amsmath}
\usepackage{amssymb}
\usepackage{tabularx, booktabs}

\usepackage{cite}
\usepackage{color}
\usepackage{hyperref}

\usepackage{lineno}

\usepackage{microtype}

\usepackage{graphicx}

\usepackage{multirow}

\usepackage[labelfont=bf,labelsep=period,justification=raggedright]{caption}

\date{}

\begin{document}

\begin{flushleft}
{\Large
\textbf{Parenclitic network analysis of methylation data for cancer identification}
}
\\
Alexander Karsakov$^{1}$,
Thomas Bartlett$^{2}$,
Artem Ryblov$^{1}$,
Iosif Meyerov$^3$,
Alexey Zaikin$^{1,2,\ast}$,
Mikhail Ivanchenko$^{1}$
\\
\bf{1}  Department of Applied Mathematics and Centre of Bioinformatics, Lobachevsky State University of Nizhny Novgorod, Nizhniy Novgorod, Russia.
\\
\bf{2} Institute for Women's Health and Department of Mathematics, University College London, London, United Kingdom.
\\
\bf{3}  Department of Mathematical Software and High-Performance Computing, Lobachevsky State University of Nizhny Novgorod, Nizhniy Novgorod, Russia.
\\
$\ast$ E-mail: alexey.zaikin@ucl.ac.uk
\end{flushleft}

\section*{Abstract}
We make use of ideas from the theory of complex networks to implement a machine learning classification of human DNA methylation data, that carry signatures of cancer development. The data were obtained from patients with various kinds of cancers and represented as parenclictic networks, wherein nodes correspond to genes, and edges are weighted according to pairwise variation from control group subjects. We demonstrate that for the $10$ types of cancer under study, it is possible to obtain a high performance of binary classification between cancer-positive and negative samples based on network measures. Remarkably, an accuracy as high as $93-99\%$ is achieved with only $12$ network topology indices, in a dramatic reduction of complexity from the original $15295$ gene methylation levels. Moreover, it was found that the parenclictic networks are scale-free in cancer-negative subjects, and deviate from the power-law node degree distribution in cancer. The node centrality ranking and arising modular structure could provide insights into the systems biology of cancer. 

\section*{Introduction}

Epigenetic information is stored in a genome in the form of heritable modifications to the chemical structure of DNA, such as methylation of CpG di-nucleotides, and a number of chemical modifications of histone proteins. It can be modulated during the lifetime of an organism by environmental signals \cite{Feinberg_2006, Cooney_2007, Christensen_2009}, and these changes persist in subsequent mitoses, as an acquired change of phenotype. However, besides the fundamental role of epigenetics in mediating environmental effects on the genome, it leaves a backdoor for environmental risk factors.

In particular, variations in DNA methylation (DNAm) accompany the early stages of human carcinogenesis \cite{Feinberg_2006}, and could, therefore, manifest as quantitative signatures of such illnesses or as a risk of their development \cite{Jaffe_2012,Hansen_2011,Teschendorff_2012a,Teschendorff_2012b}. Due to the huge number of individual CpG sites (of the order of $10^5$) at which methylation levels are assessed, there is a substantial interest in developing aggregate measures, so as to reduce the dimension of the mathematical problem, whilst still taking account of the key genomic effects of cancer.

Recently, it has been demonstrated that gene-specific measures of DNA methylation, such as mean, variance, mean derivative and suchlike reflect cancer-related changes and enable differentiation between normal and tumour tissue samples \cite{Bartlett_2013}. Furthermore, a method to construct a network representation of the data was proposed, with genes taken as nodes. Edge weights then quantify the extent to which the methylation profiles of these genes covary in the same way as the healthy population profiles do \cite{Bartlett_2014}. Some edges of this network representation were shown to be associated with survival outcome for patients with different types of cancer. It was also found that natural groupings of these prognostic edges could be identified as subnetwork modules, relevant to a number of biological functions. This indicated that epigenetic network models and measures do not just technically reduce the complexity of a computational problem, but naturally reflect intrinsic collective behaviour and interactions of such groups of genes.

Inspired by these findings, we address the problem of constructing epigenetic data networks and identifying network measures for distinguishing between normal and cancer cells. We seek a solution implementing recently developed parenclitic network analysis \cite{Zanin_2011,Zanin_2014}. This approach identifies generic biomedical measurements with nodes and specifies that an edge exists between each pair of nodes if their values for a particular subject are significantly different from the linear regression model for a control ``healthy'' group (or weight of the edge is proportional to the mismatch of the regression model). In result one obtains a network for each subject, which properties are expected to be different in health and disease.

We demonstrate that parenclictic networks built on DNAm data correctly represent source data, therefore classifiers based on routine network measures (average node degree, diameter, etc.) can produce approximately $93-99\%$ accuracy. The statistics of parenclictic networks for healthy tissues exhibits power-law tails in the node-degree distributions, that indicates substantial natural fluctuations in methylation levels. Remarkably, cancer modifications in DNAm induce a qualitative change in network topology: heavy-tailed too, they show marked deviations from power-law scaling. Exceptions are found in one case only, where the network architecture does not change noticeably and at the same time the performance of the respective classifier drops to about $90\%$.

\section*{Methods}

\subsection*{Data}

Methylation data, collected via the Illumina Infinium Human Methylation 450 platform, were downloaded from The Cancer Genome Atlas (TCGA) project \cite{Collins_2007} at level 3. Data were obtained from ten different healthy and tumour tissues: Bladder Urothelial Carcinoma (BLCA), Breast Invasive Carcinoma (BRCA), Colon Adenocarcinoma (COAD), Head and Neck Squamous Cell Carcinoma (HNSC), Kidney Renal Clear Cell Carcinoma (KIRC), Kidney Renal Papillary Cell Carcinoma (KIRP), Lung Adenocarcinoma (LUAD), Prostate Adenocarcinoma (PRAD), Thyroid Carcinoma (THCA), and Uterine Corpus Endometrioid Carcinoma (UCEC). The number of samples for each data set is shown in
Table \ref{table:1}. Raw data were pre-processed as described in \cite{Bartlett_2013}, briefly summarised as follows. First, probes were removed if they have non-unique mappings or map to SNPs (as identified in the TCGA level 3 data); probes mapping to sex chromosomes were also removed; in total 98384 probes were removed in this way from all data sets. After removal of these probes, 270985 probes with known gene annotations remained. Individually for each data set, probes were then removed if they had less than 95\% coverage across samples; probe values were also replaced if they had corresponding detection $p$-value greater than 5\%, by KNN ($k$ nearest neighbour) imputation ($k=5$).  {Methylation level has been considered across the whole gene, as a way of simplifying the large quantity of data. And whilst some information will undoubtedly be lost by rescaling all methylation values into range $[0,1]$ and characterising each gene by the average level of methylation, we still found that changes in methylation level across the whole gene were indicative of disease. In future work, it will be important to refine the analysis by considering mean methylation across specific functional regions within the gene, such as the promoter, the first exon, et cetera.} The total number of genes of interest was $15295$. 

\begin{table}[h]
\begin{center}
\begin{tabular}{|c|c|c|}

\hline
\textbf{Type of cancer} & \textbf{Number of healthy subjects} & \textbf{Number of cancer subjects} \\ \hline
BLCA            & 18                                  & 126                                \\ \hline
BRCA            & 98                                 & 586                                 \\ \hline
COAD            & 38                                 & 258                                 \\ \hline
HNSC            & 50                                  & 310                                \\ \hline
KIRC            & 160                                 & 283                                \\ \hline
KIRP            & 50                                  & 98                                 \\ \hline
LUAD            & 32                                  & 306                                \\ \hline
PRAD            & 49                                  & 176                                \\ \hline
THCA            & 50                                  & 357                                \\ \hline
UCEC            & 36                                  & 334                                \\ \hline
\end{tabular}
\caption{Number of samples obtained from normal tissue (healthy subjects) and tumour tissue (cancer subjects), see the text for abbreviations. The data were downloaded from TCGA portal.}
\label{table:1}
\end{center}
\end{table}

\subsection*{Construction of networks}

We utilize the parenclictic network approach \cite{Zanin_2011,Zanin_2014} in order to construct and analyze graphs from gene methylation data. The resulting network is a complete weighted graph: each vertex corresponds to a specific gene, and edge weight is proportional to the variation of methylation levels in specific gene pair in cancer-positive and negative phenotypes. The procedure is aimed at unveiling hidden relations between methylation levels of gene pairs and discovery of global dependencies.

The procedure originally applied to different biomedical data \cite{Zanin_2011,Zanin_2014} includes the following steps:
\begin{enumerate}
\item Select a control group from healthy tissue samples.
\item Adjust methylation levels for each pair of genes, $m_i$  and $m_j$, to a linear regression based on the control group subjects:
\begin{equation}
\label{eq:1}
m_j = \alpha_{i,j} + \beta_{i,j} m_i, 
\end{equation}
where $\alpha_{i,j}$ and $\beta_{i,j}$ are regression coefficients. 
\item Build complete weighted graph for each cancer-positive and negative sample, excluding the control group, such that each vertex corresponds to a particular gene, and edges are weighted according to
\begin{equation}
\label{eq:2}
w_{i,j} = \frac{| x_j - (\alpha_{i,j} + \beta_{i,j} x_i) |}
{\sigma_{i,j}}, 
\end{equation}
where $x_i$ and $x_j$ are respective methylation levels, and $\sigma_{i,j}$ is the standard deviation of errors in the linear regression model for control objects (\ref{eq:1}). 
\end{enumerate}

\begin{figure}[h]
{\centering\includegraphics[scale=0.35]{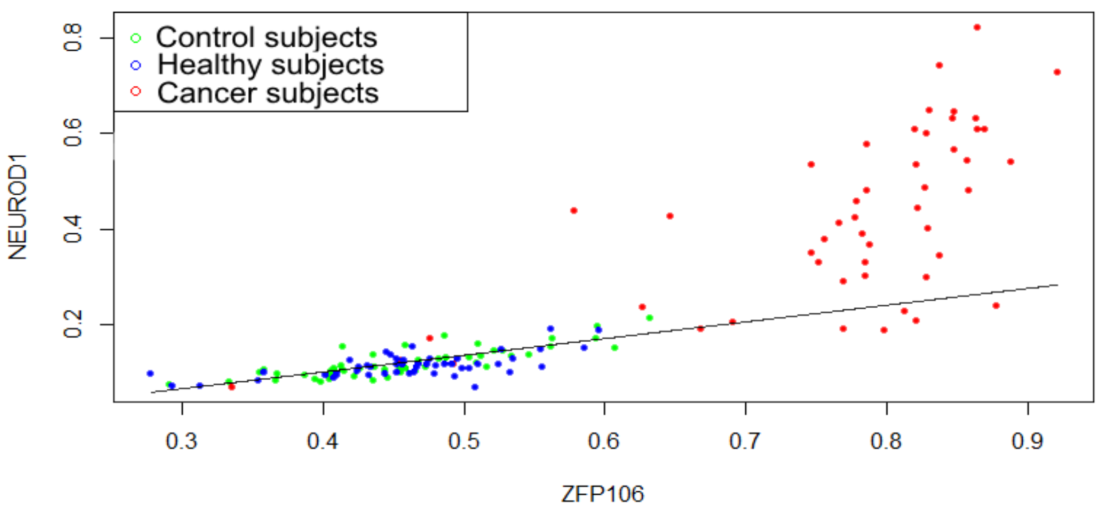}}
\caption{Determining the edge weight between ZFP106 and NEUROD1 genes.
Each point corresponds to gene methylation levels in the control group (green), other BRCA-negative (blue) and BRCA-positive (red) subjects. The solid line shows the linear regression model (\ref{eq:1}). The mismatch to (\ref{eq:2}) is, in general, a good discriminator between healthy and tumour samples. 
}
\label{fig:1}
\end{figure}

This process is illustrated in Fig. \ref{fig:1} through the use of the two genes ZFP106 and NEUROD1 for BRCA data. Each point corresponds to gene methylation levels in the control group (green), and other BRCA-negative (blue) and BRCA-positive (red) subjects. The data points for the subjects with the disease are substantially more distant from the linear regression model (solid line), as compared to the data points for the healthy ones. Thus at least some of the edge weights in the parenclictic networks for the former group will be considerably greater, potentially introducing detectable modifications in global network characteristics \cite{Zanin_2011,Zanin_2014}.  

\begin{figure}[h!]
\centering\includegraphics[scale=0.8]{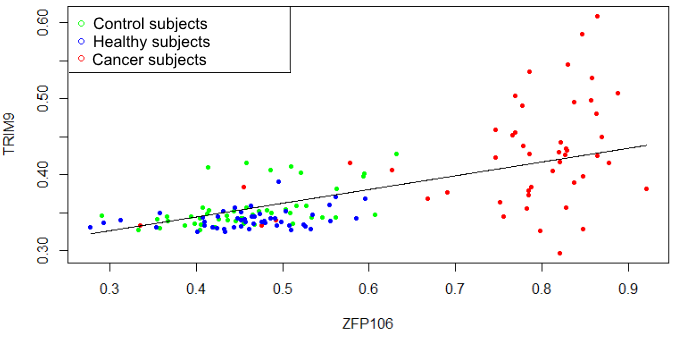}
\caption{Determining the edge weight between ZFP106 and TRIM9 genes.
Each point corresponds to gene methylation levels in the control group (green), other BRCA-negative (blue) and BRCA-positive (red) subjects. The solid line shows the linear regression model (\ref{eq:1}). While the data sets for healthy and tumour samples are quite distinct, the mismatch (\ref{eq:2}) is of the same order of magnitude for both classes. Employing the Mahalanobis distance (\ref{eq:3}) overcomes this problem.
   }
\label{fig:2}
\end{figure}

Linear regression approach is quite clear and gives good results in many cases. However, at least for data of interest, it does not always yield a decent estimation of the closeness between the control and cancer subjects. Indeed, changing one of the genes in the example above, that is, considering ZFP106 and
TRIM9 methylation levels in BRCA data, we find that all three of the clusters (control, BRCA-negative and positive) match the linear regression model (\ref{eq:1}) well (see Fig.\ref{fig:2}). Consequently, for all sample classes the ZFP106-TRIM9 edge weights (\ref{eq:2}) will, typically, be relatively small and of the same order of magnitude. At the same time, the BRCA-negative and positive clusters are visibly distinct. 

To overcome this problem we implement the Mahalanobis distance \cite{Huberty_2005}, which, essentially, measures separation between data sets. In particular, instead of (\ref{eq:2}) the edge weight $w_{i,j}$ is caclulated as:
\begin{equation}
\label{eq:3}
w_{i,j} = \sqrt{(x_{i,j} - \mu_{i,j})^T S_{i,j}^{-1} (x_{i,j} - \mu_{i,j})},
\end{equation}
where, as before, $x_i$ and $x_j$ are gene methylation levels in an investigated sample, $\mu_i$ and $\mu_j$ are the gene methylation levels in the control group, $x_{i,j} = (x_i, x_j)$, $\mu_{i,j} = (\mathbf{E}(m_i), \mathbf{E}(m_j))$, and $S_{i,j} = cov(m_i, m_j)$. In this way, the abnormal modifications in methylation of gene pairs are better captured, which, in turn, improves sample classification accuracy for our data by $1-3\%$.

\subsection*{Metrics of network topology}
The massive number of edges in constructed graphs makes a straightforward solution of machine learning classification problem intractable in practice, also manifesting a huge imbalance between the number of features and available samples. Therefore, we utilize a number of topology metrics, widely used for characterising complex networks \cite{Barabasi_2002,Boccaletti_2006,Freeman_1978}, appropriately generalised for weighted graphs $G:=(V,E)$ with $|V|$ vertices and $|E|$ edges:
\begin{itemize}
\item Node degree $deg(v)$ as the sum of incidental edges weights.
\item The distance $d(v_i,v_j)$ between the nodes $v_{i,j}\in V$, defined as the sum of the edge weights in the shortest path.
\item The diameter of the graph $G$ as the maximal distance between a pair of vertices. 
\item The degree centrality $C_D(G)$ of the graph $G$ defined as the normalised graph degree centrality $H(G)$
\begin{equation}
\label{eq:6}
C_D(G) = \frac{H(G)}{H_{max}},
\end{equation}
which is 
\begin{equation}
\label{eq:6a}
H(G) = \sum\limits_{i = 1}^{|V|}|C_D(v^*) - C_D(v_i)|,
\end{equation}
based on the node degree centrality $C_D(v) = \frac{deg(v)}{|V|}$, and where $v^*$ is the node with the maximal degree centrality, and $H_{max}=(|V|-1)(|V|-2)$ is the maximal graph degree centrality, obtained for the star topology. 

\item Graph efficiency $E_C(G)$ defined as \cite{Latora_2001}
\begin{equation}
\label{eq:8}
E_C(G) = \frac{C_C(G)}{|V|(|V|-1)},
\end{equation}
based on the graph centrality measure
\begin{equation}
\label{eq:7}
C_C(G) = \sum\limits_{i \ne j}\frac{1}{d(v_i, v_j)}.
\end{equation}

\item Betweenness centrality  $C_B(v_k)$ of a node as the number of the shortest paths the particular node $v_k$ belongs to:
\begin{equation}
\label{eq:9}
C_B(v_k) = \sum\limits_{k \ne i \ne j} \frac{\sigma_{v_i,v_j}(v_k)}{\sigma_{v_i,v_j}},
\end{equation}
where $\sigma_{v_i,v_j}$ is the number of the shortest paths between the nodes $v_i$ and $v_j$, 
among which $\sigma_{v_i,v_j}(v_k)$ passing through $v_k$.
\end{itemize}

These quantities in one way or another should reflect the expected differences between the sample classes. For instance, increasing separation of data sets produces greater edge weights and may result in substantial decrease of the graph diameter. Likewise, nodes with large centrality scores signify their key role is class distinguishing and give us a hint at biological importance.

\section*{Results and discussion}

\begin{figure}[h]
{\centering\includegraphics[scale=0.5]{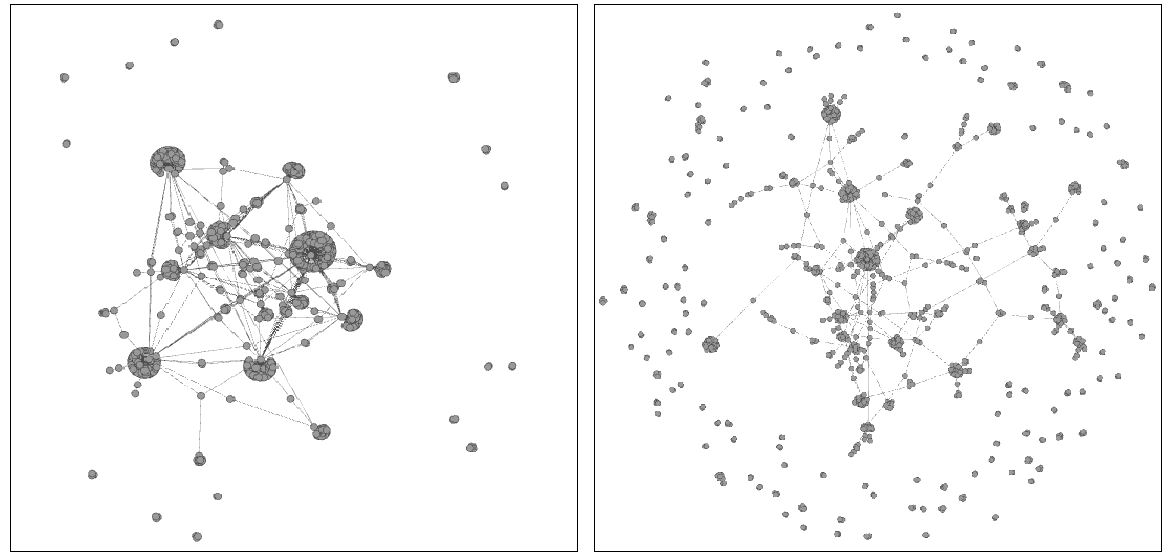}}
\caption{Typical examples of parenclictic networks constructed from gene methylation profiles for cancer (left) and normal (right) samples from BRCA data. Only a $1000$ of the strongest edges and their incident nodes are shown. Note the pronounced modular structure for the cancer network.}
\label{fig:3}
\end{figure}

As we can see from the examples of the pairwise gene methylation level diagrams, there exists intrinsic variability in both healthy and tumour tissue samples (see Figs. \ref{fig:1} and \ref{fig:2}). Therefore, parenclictic networks built from both data classes should have random-like and complex structure. To get a general idea on the appearance of the resulting networks, we present indicative examples of BRCA-positive and negative donors, plotting the $1000$ edges with largest weights and incident nodes (Fig.\ref{fig:3}). Remarkably, for cancer subjects they typically comprise several star-type subgraphs, with the most abnormally methylated genes as vertices. At the same time, healthy subjects yield considerably more homogeneous networks.  

Distinguishing between cancer-positive and negative DNAm profiles is implemented by standard machine learning classification algorithms, namely, Support Vector Machine (SVM) and Random Forest (RF) \cite{Cortes_1995,Breiman_2001}. The binary classifiers are trained on 12 topological indices calculated for parenclictic networks as described in Methods:

\begin{itemize}
\item mean, variation and maximal values of edge weights,
\item mean, variation and maximal values of vertex degree,
\item mean and variation of shortest path lengths,
\item diameter of graph,
\item degree centrality,
\item efficiency,
\item and betweenness centrality.
\end{itemize}
In addition, RF was also trained on the original gene average methylation level data. All code was written in Python 3. For classification we have used Python scikit-learn package.

In order to assess the performance of our classifiers we used the  {two-step} cross-validation technique.  {The first step included cross-validation for selecting a control group from healthy tissue samples. For data sets where the number of healthy subjects was  {less} than 50 instances, we applied 2-fold cross-validation; otherwise, 4-fold cross-validation was used. The second step involved factual 10-fold cross-validation for every data set with topology indices.}

 {Limited by the amount of data, we couldn't make use of AUC as a classification metrics. Indeed, a quarter or a half of healthy patients data were put aside after the first cross-validation step. During the 10-fold cross-validation at the second step, the data were divided further and, after all, each fold contained one or two healthy patients at best. In result, AUC could be calculated in rare cases only.}

\newcolumntype{Y}{>{\centering\arraybackslash}X}
\begin{table}[h]
\begin{tabularx}{\textwidth}{|c *{6}{|Y}|}

\cline{2-7}
\multicolumn{1}{c|}{} 
 & \multicolumn{3}{c|}{{ Classification with topology indices}}  
 & \multicolumn{3}{c|}{{ Classification with gene methylation levels}}\\
\hline
 Cancer & Accuracy & Specificity & Sensitivity & Accuracy & Specificity & Sensitivity\\
\hline
BLCA & 95.89\% & 77.77\% & 99.19\%
 & 95.95\% & 66.66\% & 98.38\%\\ \hline
BRCA & 96.96\% & 91.87\% & 98.11\%
 & 97.82\% & 87.67\% & 98.96\%\\ \hline
COAD & 99.30\% & 94.73\% & 100.00\%
 & 99.33\% & 94.73\% & 100.00\%\\ \hline
HNSC & 96.70\% & 85.57\% & 98.69\%
 & 98.60\% & 92.27\% & 99.36\%\\ \hline
KIRC & 98.63\% & 98.75\% & 98.92\%
 & 98.90\% & 98.62\% & 98.14\%\\ \hline
KIRP & 99.17\% & 97.72\% & 100.00\%
 & 96.03\% & 95.89\% & 95.80\%\\ \hline
LUAD & 99.43\% & 93.75\% & 99.01\%
 & 99.39\% & 93.72\% & 100.00\%\\ \hline
PRAD & 90.40\% & 71.58\% & 89.65\%
 & 92.58\% & 85.12\% & 94.76\%\\ \hline
THCA & 93.12\% & 70.03\% & 96.60\%
 & 94.33\% & 71.90\% & 97.40\%\\ \hline
UCEC & 98.62\% & 91.67\% & 99.10\%
 & 99.20\% & 91.67\% & 100.00\%\\ \hline
\end{tabularx}

\caption{Classification accuracy of RF machine learning algorithm trained on topology indices of parenclictic networks, along with the RF performance on the original gene average methylation level data for different types of cancer.}
\label{table:2}
\end{table}

The results, summarized in Table \ref{table:2}  {and Table \ref{table:3}}, demonstrate an excellent performance of classifiers trained on network measures for almost all kinds of cancer, with accuracy in the range $93- {99}\%$. Interestingly, the performance does not manifest any considerable dependence on the type of classification algorithm in most cases, with the only noticeable  {dissimilarity} of RF observed for  {BRCA, PRAD and THCA} groups. We suggest that the accuracy of classification would not depend strongly on the particular choice of machine learning algorithm.
\newcolumntype{Y}{>{\centering\arraybackslash}X}
\begin{table}[h]
\begin{tabularx}{1.0\textwidth}{|c *{6}{|Y}|}

\cline{2-7}
\multicolumn{1}{c|}{} 
 & \multicolumn{3}{c|}{ {Classification with topology indices}}  
 & \multicolumn{3}{c|}{ {Classification with gene methylation}}\\
\hline
 Cancer & Accuracy & Specificity & Sensitivity & Accuracy & Specificity & Sensitivity\\
\hline
BLCA & 95.89\% & 77.78\% & 98.39\%
 & 95.95\% & 66.66\% & 98.38\%\\ \hline
BRCA & 93.62\% & 83.33\% & 95.21\%
 & 97.82\% & 87.67\% & 98.96\%\\ \hline
COAD & 99.33\% & 94.74\% & 100.00\%
 & 99.33\% & 94.73\% & 100.00\%\\ \hline
HNSC & 95.89\% & 84.62\% & 97.40\%
 & 98.60\% & 92.27\% & 99.36\%\\ \hline
KIRC & 96.36\% & 95.00\% & 97.14\%
 & 98.90\% & 98.62\% & 98.14\%\\ \hline
KIRP & 98.75\% & 100.00\% & 97.62\%
 & 96.03\% & 95.89\% & 95.80\%\\ \hline
LUAD & 99.41\% & 93.75\% & 100.00\%
 & 99.39\% & 93.72\% & 100.00\%\\ \hline
PRAD & 87.40\% & 61.33\% & 94.82\%
 & 92.58\% & 85.12\% & 94.76\%\\ \hline
THCA & 96.09\% & 84.62\% & 97.73\%
 & 94.33\% & 71.90\% & 97.40\%\\ \hline
UCEC & 98.39\% & 88.89\% & 99.40\%
 & 99.20\% & 91.67\% & 100.00\%\\ \hline
\end{tabularx}
\caption{Classification accuracy of SVM algorithm trained on topology indices of parenclictic networks, along with the RF performance on the original gene average methylation level data for different types of cancer.}
\label{table:3}
\end{table}

Comparing performance of the RF classifier built on the network measures and the original average methylation levels of 15295 genes, we do not notice systematic weaknesses of the latter. Moreover, in some cases, using network measures enhances classification results. Except for the case with  {PRAD}, which we will discuss hereinafter, the network measures seem to incorporate cancer modifications of gene methylation levels very well, and the substantial reduction of the complexity of classification task does not impair the resulting accuracy. Another benefit of implementing classification on network measures is that in this case the number of features is considerably smaller than the number of samples, which prevents overfitting.

\begin{figure}[h]
{\centering\includegraphics[scale=0.95]{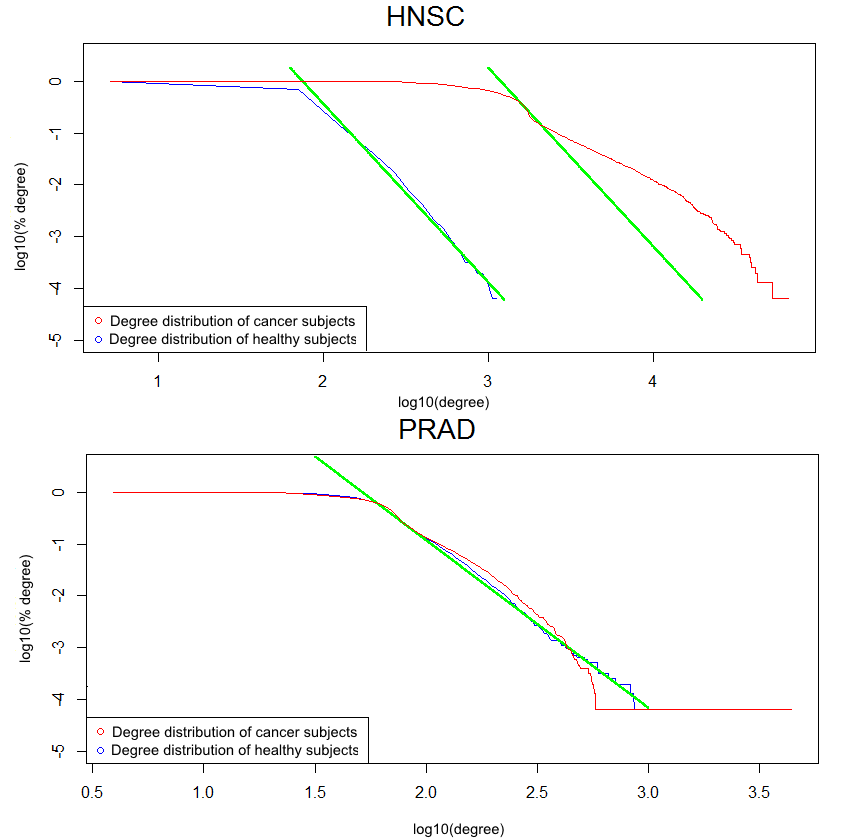}}
\caption{Average complementary cumulative node degree distribution (the fraction of nodes, for which the degree exceeds a given value) for HNSC (top) and PRAD (bottom) subjects. Green lines on the log-log plots display the power-law model best fit.}
\label{fig:4}
\end{figure}

Let us investigate it in more detail, why the performance of the network measure classifiers for  {PRAD} is  {slightly} poorer, at about $ {87-90}\%$. Clearly, the answer must be sought in the less pronounced differences between the classes in the global network characteristics. To get a deeper insight we analysed the node degree distributions for parenclictic networks constructed from the data, corresponding to different types of cancer. Remarkably, the results consistently revealed that for cancer-negative subjects the networks are scale-free, with the complementary cumulative degree distributions closely following a power-law (see Fig.\ref{fig:4} for representative cases). On the contrary, parenclictic networks for cancer-positive subjects demonstrate pronounced deviations from power-law scaling, except for PRAD, where the node degree statistics does not change significantly and network measures give worse classification accuracy (Fig.\ref{fig:4}). This suggests that the further work here should primarily focus on modifying the network construction method itself, rather than introducing other network measures. 

\section*{Conclusions}

We demonstrated that the parenclictic network approach can be successfully implemented in order to obtain graph representations of gene methylation levels for cancer-positive and negative subjects. These graphs can be characterised by 12 global network measures, which provide a basis for binary classification by routine machine learning algorithms, Support Vector Machine and Random Forest. For almost all cancer types the performance of both algorithms remains much the same so that the particular choice does not seem crucial. 

Comparing to performance of Random Forest classifier built on the original average methylation levels of 15295 genes does not reveal a substantial difference, with the accuracy reaching $90- {99}\%$ for  {all} the types of cancer. This means that a cardinal reduction to 12 features, topological indices, does not lead to the loss of important information. Yet another strong benefit of the considerably decreased complexity is the avoidance of overfitting, which could be a serious problem when using the original 15295 gene methylation levels as features, due to the relatively small number of samples.

Finally, network analysis does not only allow identifying cancer methylation profiles but provides additional insight into the systems biology of cancer. That is, the nodes with high centrality are good candidates for playing a significant role in cancer development. Strong deviations from scale-free node degree distributions for almost all cancer types is a hallmark of the global changes in the network topology induced by cancer, which remains to be understood. Another open question is the modular properties of the arising parenclictic networks, observed by naked eye inspection, and their correspondence to biological functions.

\section*{Acknowledgments}
The Authors acknowledge Russian Foundation for Basic Research, grant No. 14-04-01202 (A.R., I.M., A.Z.).
 {M.I. acknowledges support of Russian Science Foundation grant No.\ 14-12-00811 (network statistics analysis).
Computations were carried out on the Lobachevsky University supercomputer.}


\end{document}